\documentclass[final,5p,times,twocolumn]{elsarticle} 

\usepackage{amssymb,color}

\newcommand{\red}{\color{red}}
\newcommand{\black}{\color{black}}
\renewcommand{\red}{\black}

\journal{Physics Letters B}

\begin{document}

\begin{frontmatter}



\title{On the impact of magnetic-field models in galaxy clusters on
constraints on axion-like particles from the lack of irregularities in
high-energy spectra of astrophysical sources}


\author{Maxim Libanov$^{1,2}$ and Sergey Troitsky$^{1}$}

\address{$^{1}$ Institute for Nuclear
Research of the Russian Academy of Sciences,\\
60th October Anniversary
Prospect 7a, Moscow 117312, Russia\\
$^{2}$ Moscow Institute of Physics and Technology,\\
Institutskii per., 9, 141700, Dolgoprudny, Moscow Region, Russia}

\begin{abstract}
Photons may convert to axion-like particles (ALPs) in external magnetic
fields. Under certain conditions, this effect should result in
irregular features in observed spectra of astrophysical sources. Lack of
such irregularities in particular spectra was used to constrain ALP
parameters, with two most popular sources being the radio galaxy NGC~1275
and the blazar PKS~2155$-$304.  The effect and, consequently, the
constraints, depend on the magnetic fields through which the light from
the source is propagated. Here, we revisit ALP constraints from gamma-ray
observations of NGC~1275 taking into account the regular magnetic field of
the X-ray cavity observed around this radio galaxy. This field was not
accounted for in previous studies, which assumed a model of purely
turbulent fields with coherence length much smaller than the cavity size.
For the purely regular field, ALP constraints are relaxed considerably,
compared to the purely turbulent one.
While the actual magnetic
field around a source is an unknown sum of the turbulent and ordered
components, the difference in results gives an estimate of the theoretical
uncertainty of the study and calls for detailed measurements of magnetic
fields around sources used to constrain ALP properties in this approach.
\end{abstract}




\end{frontmatter}


\section{Introduction}
\label{sec:intro}
Pseudoscalar axion-like particles (ALPs; for a review, see e.g.\
Ref.~\cite{Ringwald-rev}) are common in various extensions of the Standard
Model of particle physics. Their defining property is the
interaction with the electromagnetic field, allowing in particular for
photon/ALP mixing in the external magnetic field \cite{RaffStod}.
Experimentally allowed values of the photon-ALP coupling are however very
low, which makes large-scale astrophysical environments a proper place to
search for manifestations of this interaction. The astrophysical searches
include studies of stellar energy losses, where indications to
losses larger than predicted may be explained by interactions of
ALPs, see Ref.~\cite{Losses} for a review. In particular, cooling of
helium-burning stars~\cite{HBstars} may suggest ALP-photon couplings just
below the present limits from the solar ALP searches by the CERN Axion
Solar Telescope, CAST \cite{CAST}. The same range of couplings is favoured
by ALP explanations of possible ``anomalous transparency''
effects~\cite{hardening, 3c279, 2photons, HornsMeyer, gamma} in the
propagation of energetic gamma rays from distant sources~\cite{Csaba,
DARMA, Serpico, FRT, 1207.0776clusters, Meyer-evidence, Galanti-spindex}
(for a short review, see Ref.~\cite{ST-rev}). These studies are however
very sensitive to measurements of redshifts of the emitting sources, so
the significance of the observed effects remains uncertain \cite{gamma2}
and other ways to explore the relevant part of the ALP parameter space are
welcome.

Under certain conditions, photon-ALP mixing may result in oscillatory
features in the spectra of astrophysical photon sources seen through
regions filled with magnetic fields \cite{astro-ph/0410501}. While precise
shape of these irregularities and the photon energies at which they appear
depend on the ALP parameters and magnetic-field configurations, these
features are not expected in (otherwise smooth) astrophysical spectra.
Lack of these irregularities in observed spectra might in principle be
used to constrain ALP parameters, see e.g.\
Refs.~\cite{0806.0411, 1205.6428, 1305.2114, 1406.5972, 1804.09443,
1811.03548}.

Several attempts to follow this way have been made in recent years,
exploiting the observed smoothness of X- and gamma-ray spectra of strong
emitters presumably embedded in astrophysical magnetic fields. In
particular, these included analyses of gamma-ray observations of
PKS~2155$-$304 \cite{1311.3148, 1802.08420, 1906.00357}, NGC~1275
\cite{1603.06978, 1805.04388} and several Galactic sources
\cite{1801.01646, 1801.08813, 1804.07186} relevant for ALP masses $\sim
\left(10^{-9} \, - \, 10^{-7}\right)$~eV (overlapping with the range
invoked for explanations of the ``anomalous transparency'' effects). For
lower ALP masses, X-ray observations of 3C~218 \cite{1304.0989}, the same
NGC~1275 \cite{1605.01043, 1712.08313, 1907.05475}, M~87 \cite{1703.07354}
and some other extragalactic sources \cite{1704.05256} were used.
Refs.~\cite{1801.01646, 1801.08813} found some favorable ranges in the ALP
parameter space, though with ALP-photon couplings above the 95\%~CL CAST
upper limit. Other works put limits on the ALP parameters which are often
quoted on equal footing with other astropysical and laboratory constraints
\cite{PDG}. The aim of the present work is to demonstrate how much these
constraints depend on the assumptions about magnetic fields involved and
how much they could change if different astrophysically motivated field
configurations are assumed.

As an example, we concentrate on the case of NGC~1275, the central radio
galaxy of the Perseus cluster and the most popular target for the searches
for spectral irregularities in the ALP context \cite{1603.06978,
1805.04388, 1605.01043, 1712.08313}. We briefly review these studies in
Sec.~\ref{sec:previous} and emphasize that they use a theoretical model of
the magnetic field in the Perseus cluster, which is based on observations
of other clusters, because of the lack of relevant observations in
Perseus. This model assumes purely turbulent magnetic fields. In the
present study, we revisit these constraints taking into account
large-scale ordered magnetic fields which are expected to be present
around giant radio galaxies from the interaction between lobes and
environment in clusters \cite{a-p/0204443, 1108.0430, 1203.4582,
1811.06266, 1812.07900}. Evidence for the regular magnetic fields ordered
at scales $\gtrsim 100$~kpc have been found in several galaxy clusters
\cite{1101.1807, 1612.01764}, see also reviews \cite{1205.1919,
Han-ARAA55(17)111}. Possible presence of the regular magnetic fields
around radio galaxies is crucial for the present consideration.

In Sec.~\ref{sec:field}, we concentrate on the magnetic field in the
Perseus cluster. Only two relevant observations of the Faraday rotation
have been reported in the literature: Ref.~\cite{Taylor-FRM} gives a
measurement at the very center of the cluster, used in previous studies
\cite{1603.06978, 1805.04388, 1605.01043, 1712.08313} to normalize the
field strength at this point; Ref.~\cite{FRM-map} presents a map of
rotation measurements across the cluster (not used in previous studies).
We recall numerous radio and X-ray observations indicating
the presence of an X-ray cavity, or a radio mini-halo, around NGC~1275,
consistent with the lobe-environment interaction \cite{MNRAS264(93)L25,
a-p/0207290, a-p/0503318, 1701.03791}. An analytical model of the regular
magnetic field in such a cavity was presented (and supported by numerical
simulations) in Ref.~\cite{1008.5353}; see also Refs.~\cite{1011.0030,
1107.2599} and especially \cite{1108.3344} for more detailed simulations.
We use this analytical model as a proxy to the regular field in the
central X-ray cavity of the Perseus cluster, consistent with rotation
measurements of Refs.~\cite{Taylor-FRM, FRM-map}, as well as with X-ray
observations of large-scale structures in Perseus \cite{Churazov,
1810.07380} and corresponding simulations \cite{1506.06429}.

Next, in Sec.~\ref{sec:NGC1275}, we turn to the effect of the ordered
magnetic field on the observed photon spectrum, and consequently on
constraints on the ALP parameters. As an example, we readdress the case of
NGC~1275 and Fermi-LAT data~\cite{1603.06978}.
In general, magnetic field in the cluster is the sum of regular and
turbulent components, and their various combinations can fit scarce
observational data. Since both components contribute to the
observed Faraday rotation, stronger regular field
implies weaker turbulent one. In particular, previous studies assumed
purely turbulent models; here we consider the opposite case and assume a
purely regular field, which is described in Sec.~\ref{sec:field} and
agrees with observations. Comparison of the results for these two limiting
cases gives an estimate of the theoretical uncertainty of the constraints
obtained in this way. We take the spectrum used in Ref.~\cite{1603.06978}
and demonstrate that, for the assumed purely regular field configuration,
the fits with and without ALPs are equally good for a large part of the
parameter space, so that the resulting constraints on ALP parameters are
much weaker than those obtained in Ref.~\cite{1603.06978} for the opposite
case of purely turbulent field models. This is not surprising because
spectral irregularities are enhanced when the field extends to the scales
much larger than its coherence length, see e.g.\ Ref.~\cite{1311.3148}.
Therefore, we expect a similar effect on other constraints obtained in the
same way.

%
In
Sec.~\ref{sec:concl}, we briefly reiterate our main conclusion:
constraints on the ALP parameters from the lack of irregularities
in the spectra of astrophysical sources
are very sensitive to the assumptions about magnetic fields surrounding
the sources. In particular, regular fields, which are expected to be
present close to these strong active galaxies, are very important.
Detailed studies of the magnetic fields are required to reduce the
uncertainties and to obtain firm limits on ALP parameters in this way.

\section{Previous studies of NGC~1275 in the ALP context.}
\label{sec:previous}
NGC~1275 is a radio galaxy located in the center of the Perseus
cluster. It is a bright X-ray and gamma-ray source. Thanks to the
intensity of the source, sufficient statistics could be collected to
obtain detailed spectra suitable for the search for spectral features at
high confidence level, which justified the choice of this source as a
target for several attempts to constrain ALP parameters from the lack of
irregularities in the spectrum. In Ref.~\cite{1603.06978}, Fermi-LAT
collaboration analyses their data in great detail, develops a statistical
procedure and presents 95\% CL exclusion limits for ALP parameters.
Authors of Ref.~\cite{1805.04388} supplement Fermi-LAT data by published
results of the MAGIC observations of the same target \cite{1602.03099} to
extend the energy range covered by the spectrum and, consequently, the
excluded region in the ALP parameter space. Refs.~\cite{1605.01043,
1712.08313} used X-ray spectra of NGC~1275 obtained by Chandra to
constrain ALPs with lower masses. In what follows, we will mainly refer to
the Fermi-LAT study~\cite{1603.06978} as an example.

Measurements of astrophysical magnetic fields on galactic and larger
scales often rely on the Faraday rotation of the polarization
plane~\cite{1205.1919, Han-ARAA55(17)111}. The approach is not easy to
implement directly since it requires sources of polarized emission behind
the magnetic-field region, observed at different wavelengths $\lambda$ so
that the typical $\lambda^{2}$ dependence may be traced in the
polarization angle. In addition, the outcome is the product of the
longuitudinal component of the magnetic field and the electron density,
integrated over the line of sight, so some additional information and/or
theoretical input are required to reconstruct the magnetic-field structure
and values; note in particular that the other, transverse component of the
field is relevant for the ALP-photon mixing. Still, Faraday rotation
measurements remain the best available mean to reconstruct magnetic fields
in galaxy clusters. Sadly, these measurements for the Perseus cluster are
very limited and do not allow to uniquely reconstruct the field from
observations. That is why a theoretical model of the magnetic field in the
cluster was used in Refs.~\cite{1603.06978, 1805.04388, 1605.01043,
1712.08313}, and the real measurement in only one direction, towards the
very center of the cluster~\cite{Taylor-FRM}, was used to construct the
model.

Following general considerations of Ref.~\cite{1406.5972},
Ref.~\cite{1603.06978} models the magnetic field in the Perseus cluster as
purely turbulent, with the maximal coherence length of 35~kpc. The model
used there is motivated by observations of clusters other than Perseus
since, for Perseus, no detailed data are available. The model has 6
parameters, and reported ALP constraints correspond to their fiducial
values. While the effect of variations of these parameters, one by one, on
the resulting likel{\red i}hood of the spectral fit was studied,
see Fig.~7 of the Supplemental Material of
Ref.~\cite{1603.06978}, none of the considered variations
included a regular component. However, if a regular field component
exists, it contributes to the rotation measurement \cite{Taylor-FRM} so
that the amplitude of the turbulent component, and consequently the
strength of possible ALP-induced spectral irregularities, is smaller.

\section{Regular magnetic fields in the Perseus cluster}
\label{sec:field}
In the center of the Perseus cluster, like in many other clusters
containing a large active galaxy, an X-ray cavity was observed, spatially
coinciding with the so-called radio mini-halo~\cite{MNRAS264(93)L25,
a-p/0207290, a-p/0503318, 1701.03791}. This cavity most probably is
a result of the interaction between outflows of NGC~1275 and the
intracluster gas. The X-ray cavity size, according to Chandra
observations~\cite{a-p/0503318}, is 93~kpc, while the radio mini-halo
extends slightly further~\cite{1701.03791}. Modelling~\cite{1506.06429} of
the slashing cold front in the Perseus cluster, see e.g.\
Ref.~\cite{1810.07380}, indicates that the X-ray cavity should be filled
with relatively high magnetic field to support the required pressure. This
magnetic-field region is seen on the rotation-measure map of the Perseus
cluster~\cite{FRM-map} and explains high values of the rotation measure
observed in Ref.~\cite{Taylor-FRM} for the very center of the cluster.

\begin{figure}
\centerline{\includegraphics[width=0.95\columnwidth]{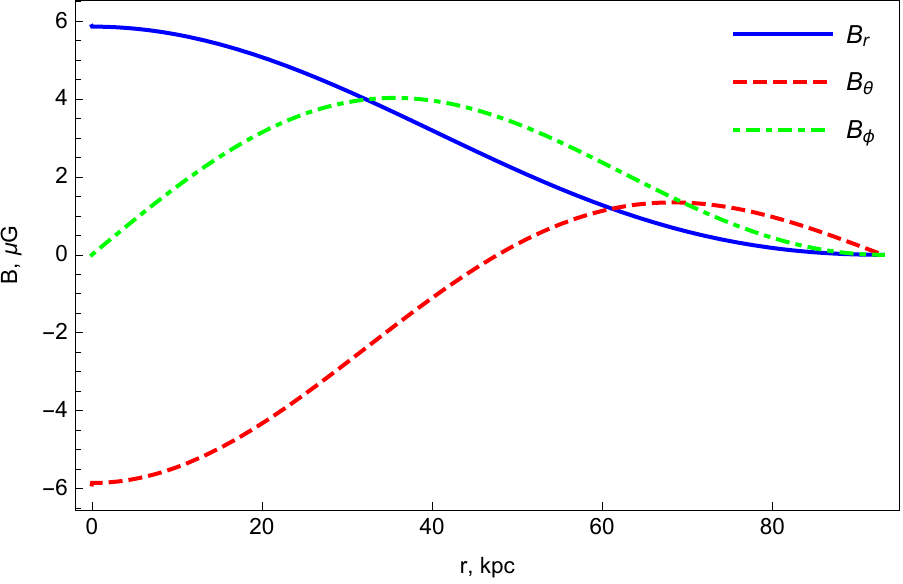}}
\caption{\label{fig:field}
Longuitudinal, $B_{r}$, and two transverse, $B_{\theta}$ and $B_{\phi}$,
components of the magnetic field solution~\cite{1008.5353} for the X-ray
cavity around NGC~1275.}
\end{figure}

For the X-ray cavities, blown by radio-galaxy jets in the intracluster
plasma, magnetic fields are expected to be regular at large scales.
Ref.~\cite{1008.5353} suggests the following consistent magnetic-field
solution,
\[
B_{r}=2\cos\theta f(r_{1})/r_{1}^{2},
\]
\[
B_{\theta}=-\sin\theta f'(r_{1})/r_{1},
\]
\[
B_{\phi}=\alpha\sin\theta f(r_{1})/r_{1},
\]
where
\[
f=C
\left(\alpha \cos(\alpha r_{1})- \sin(\alpha r_{1})/r_{1} \right) - F_{0}
r_{1}^{2}/\alpha^{2},
\]
\[
F_{0}=C \alpha^{2} \left(\alpha \cos\alpha - \sin\alpha   \right),
\]
$\alpha$ is the lowest nonzero root of
$\tan\alpha=3\alpha/(3-\alpha^{2})$, $r_{1}\equiv r/R$ for the cavity
radius $R$ and $C$ is the normalization constant determined by the field
value at $r=0$. This analytical solution is supported also by numerical
simulations in Ref.~\cite{1008.5353}. In the next section, we use this
solution as a model of the regular field in the X-ray cavity
around NGC~1275, assuming the viewing angle $\theta=45^{\circ}$ and the
cavity radius of $R=93$~kpc~\cite{a-p/0503318}. The normalization of the
field is chosen in such a way that the Faraday Rotation Measurement of
$\sim 7300$~rad/m$^{2}$ \cite{Taylor-FRM} for the central direction is
reproduced for the electron density derived from X-ray
observations~\cite{Churazov}. This normalization gives the total field in
the center of 8.3~$\mu$G. The field components for the chosen line of sight
are plotted in Fig.~\ref{fig:field}.

We should note that the central X-ray cavity may be not the only place in
the Perseus cluster where fields ordered at large scales are present.
Ref.~\cite{FRM-map} discusses also an organised structure in the
rotation-measure map on the $\sim$Mpc scales, possibly associated
\cite{AN-327-545} with a shock caused by the interaction of the
intracluster matter with intergalactic matter in the large-scale structure
filament, which may or may not be similar to the regular field structures
observed in other galaxy clusters~\cite{1101.1807, 1612.01764} under
similar conditions. In addition, X-ray observations reveal~\cite{Churazov,
1810.07380} large-scale ($>100$~kpc) spiral structure around NGC~1275,
which also may host regular magnetic fields, like it happens in our own
Galaxy. We stress therefore that the structure of regular magnetic fields
in the Perseus cluster is rich and unknown, lacking explicit measurements.
In the example in the next section, we concentrate on the best-studied
X-ray cavity field only.

\section{Example: ALP constraints from the Fermi-LAT
spectrum of NGC~1275 for the X-ray cavity magnetic field}
\label{sec:NGC1275}
Consider now the interaction of the electromagnetic field $A_{\mu}$ and
the ALP field $a$ described by the following Lagrangian,
\[
\mathcal{L} = -\frac{1}{4}F_{\mu\nu}F^{\mu\nu}+\frac{1}{2}\left(\partial a
\right)^{2}-\frac{1}{2}m^{2}a^{2}-\frac{1}{4}ga F_{\mu\nu}\tilde
F^{\mu\nu},
\]
where $F_{\mu\nu}$ is the electromagnetic stress tensor, $\tilde
F_{\mu\nu}=\frac{1}{2}\epsilon_{\mu\nu\rho\lambda}F^{\rho\lambda}$, $m$ is
the ALP mass and $g$ is the ALP-photon coupling constant. The last term is
responsible for axion-photon mixing in the external magnetic
field~\cite{RaffStod}, which in the adiabatic approximation is
most conveniently described in terms of the density matrix $\rho$ obeying
the equation
\begin{equation}
\!\!\!\!\!\!\!
\!\!
 i \frac{d\rho(y)}{dy}=\left[\rho(y),\mathcal{M}(E,y)  \right],
~~~
\mathcal{ M}\!=\!\frac{1}{2}\!\left(
\!
\begin{array}{ccc}
0 & 0 & -igB_{\theta}\\
0 & 0 & -igB_{\phi}\\
igB_{\theta} & igB_{\phi} & \frac{m^{2}}{E}
\end{array}
 \!  \right).
\label{Eq:M}
\end{equation}
For an initial unpolarized purely photon state,
\begin{equation}
\rho(0)=\mbox{diag}\left(1/2,1/2,0\right).
\label{Eq:rho0}
\end{equation}
The probability to observe an
unpolarized photon at a distance $y$ from the source is given by the sum
$\rho_{11}(y)+\rho_{22}(y)$ of the components of the solution $\rho(y)$ to
Eq.~(\ref{Eq:M}) with the boundary condition (\ref{Eq:rho0}).
Here, we assumed that the mixture of two transverse states of
photons and one state of ALP is propagating along the direction $y$ with
the energy $E$, and kept only terms relevant for the present study.
The full expression for the matrix $\mathcal{ M}$ can be found e.g.\ in
Ref.~\cite{FRT}, but entries not shown in Eq.~(\ref{Eq:M}) can safely be
neglected for energies, fields and distances studied here.

\begin{figure}
\centerline{\includegraphics[width=0.95\columnwidth]{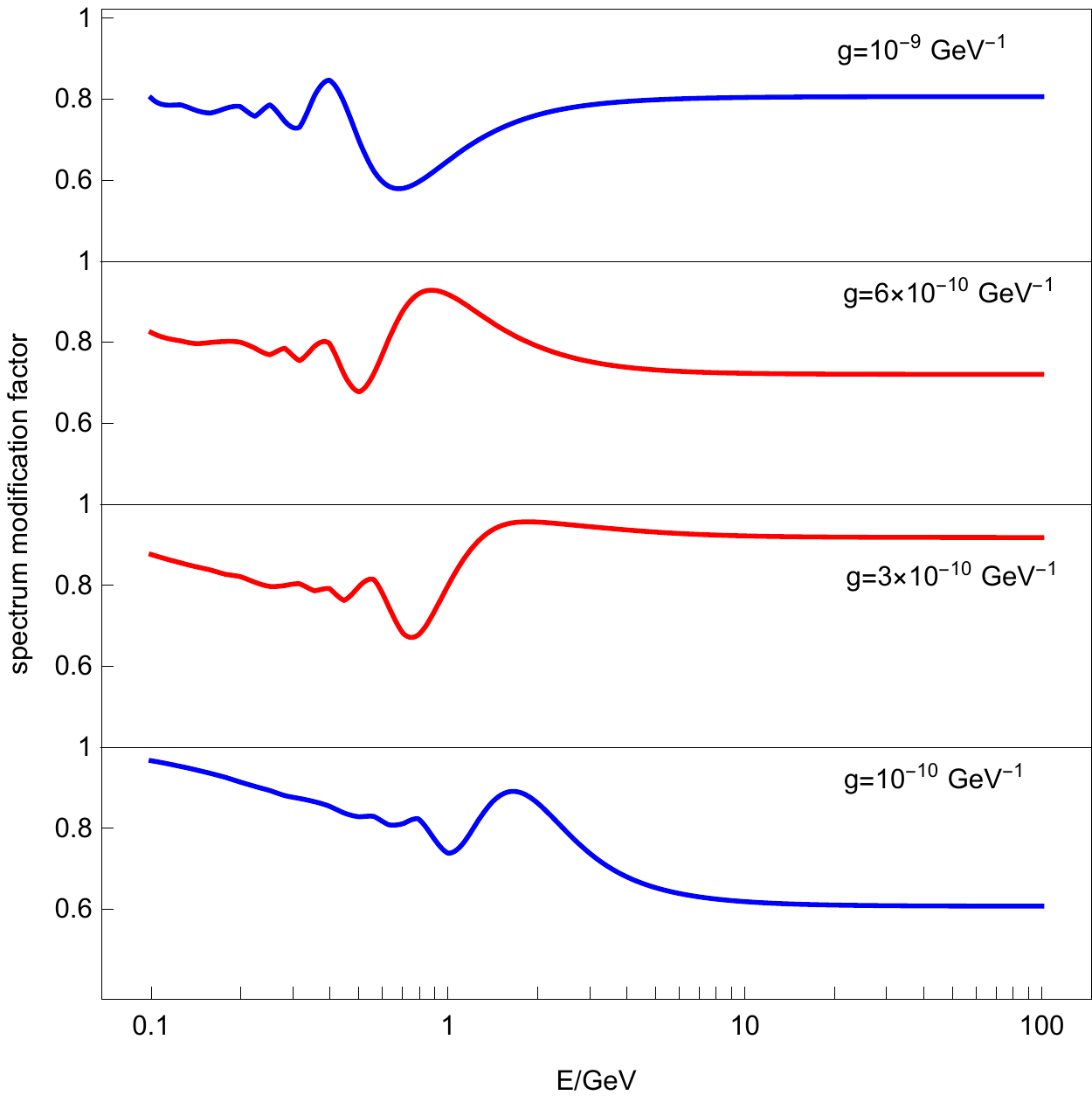}}
\caption{\label{fig:pconv}
Modification factor for the photon spectrum of NGC~1275: the probability
to detect a photon with energy $E$ from the source located in the center of
the X-ray cavity magnetic field in the presence of ALP convolved with the
Fermi-LAT energy resolution. The ALP mass $m=2$~neV and four different
values of $g$, shown in the plots, are assumed. For two
values of $g$ (red curves), stronger irregularities are present in
the most constraining region around $\sim 1$~GeV, and these values are
excluded at the 95\% confidence level. }
\end{figure}

We use the magnetic-field model for the X-ray cavity around
NGC~1275 described in Sec.~\ref{sec:field}. We assume that effects of the
turbulent field in the outskirts of NGC~1275 are subleading, like it is
normally assumed for our own Galaxy\red\footnote{\red We tested
explicitly that the addition of turbulent magnetic field, modelled
as in Ref.~\cite{1603.06978}, having the mean amplitude of
1~$\mu$G at the boundary of the X-ray cavity and extending up to
50~kpc from the center, does not change the modification factor by
more than $2\dots 3\%$, which is within statistical uncertainties
of the NGC~1275 spectral data points.}\black. For the Milky-Way magnetic
field, we use the BSS model of Pshirkov et al.\ \cite{PshirkovGMF}. The
need for a choice of the Galactic field model introduces additional
systematic uncertainties to the results. The model we use is conservative
in the sense that it predicts smaller photon-ALP oscillations
probabilities for sources away from the Galactic plane (the Galactic
latitude of NGC~1275 is $b\approx -13^{\circ}$ and this difference is not
dramatic). We assume the symmetry axis of the cavity field to be
co-oriented with the observed jets, position angle $\approx 147^{\circ}$
(from North to East) in Galactic coordinates. The redshift of NGC~1275 is
only $z=0.018$, so effects of pair production on the extragalactic
background light are negligible at the Fermi-LAT energies. We assume a
source of unpolarized photons in the center of the X-ray cavity and solve
Eqs.~(\ref{Eq:M}), (\ref{Eq:rho0}) numerically to obtain the probability
to observe photons with a given energy at the Earth. Then, we concentrate
on the Fermi-LAT data for the EDISP3 event type shown in
Ref.~\cite{1603.06978}. We convolve the probability with the
energy-dependent instrumental energy resolution presented in Supplemental
Material of Ref.~\cite{1603.06978} to obtain the ratio of the observed
photon spectrum to the emitted one, which we call the spectrum
modification factor. For illustration, it is presented in
Fig.~\ref{fig:pconv}
for benchmark values of ALP parameters.

Next, we follow the logic of Ref.~\cite{1603.06978} and find best-fit
spectra with and without ALPs and compare the quality of the fits. We use
the EDISP3 event type spectrum of NGC~1275 from Fig.~1 of
Ref.~\cite{1603.06978}. While the actual emitted spectrum is unknown, most
of active-galaxy gamma-ray fluxes $F(E)$ are perfectly fit by the so-called
log-parabola function,
\begin{equation}
F(E)=F_{0} \left(E/E_{0}\right)^{-\left(\alpha + \beta \log(E/E_{0})
\right)}
,
\label{Eq:logpar}
\end{equation}
where, following Ref.~\cite{1603.06978}, we fix $E_{0}=0.53$~GeV and keep
three parameters, $F_{0}$, $\alpha$ and $\beta$, free. We then perform two
independent fits of the data, each with the free parameters: one by the
log-parabola spectrum (\ref{Eq:logpar}) and another by the same function
multiplied by the spectrum modification factor for given values of the ALP
parameters. The fits are performed by the usual
chi-square minimization adopted to account for asymmetric statistical
errors \cite{physics/0401042}. Upper limits are treated as zero values
with the corresponding errors. This is a simplification compared to the
procedure of Ref.~\cite{1603.06978}, where the full likel{\red i}hood
function based on non-Gaussian distribution of errors was used; however,
this is sufficient for our purposes since we are interested in the best
fit only\footnote{\red We checked that for the fit without ALPs, our
results agree with those presented in Ref.~\cite{1603.06978}.}.

\begin{figure}
\centerline{\includegraphics[width=0.95\columnwidth]{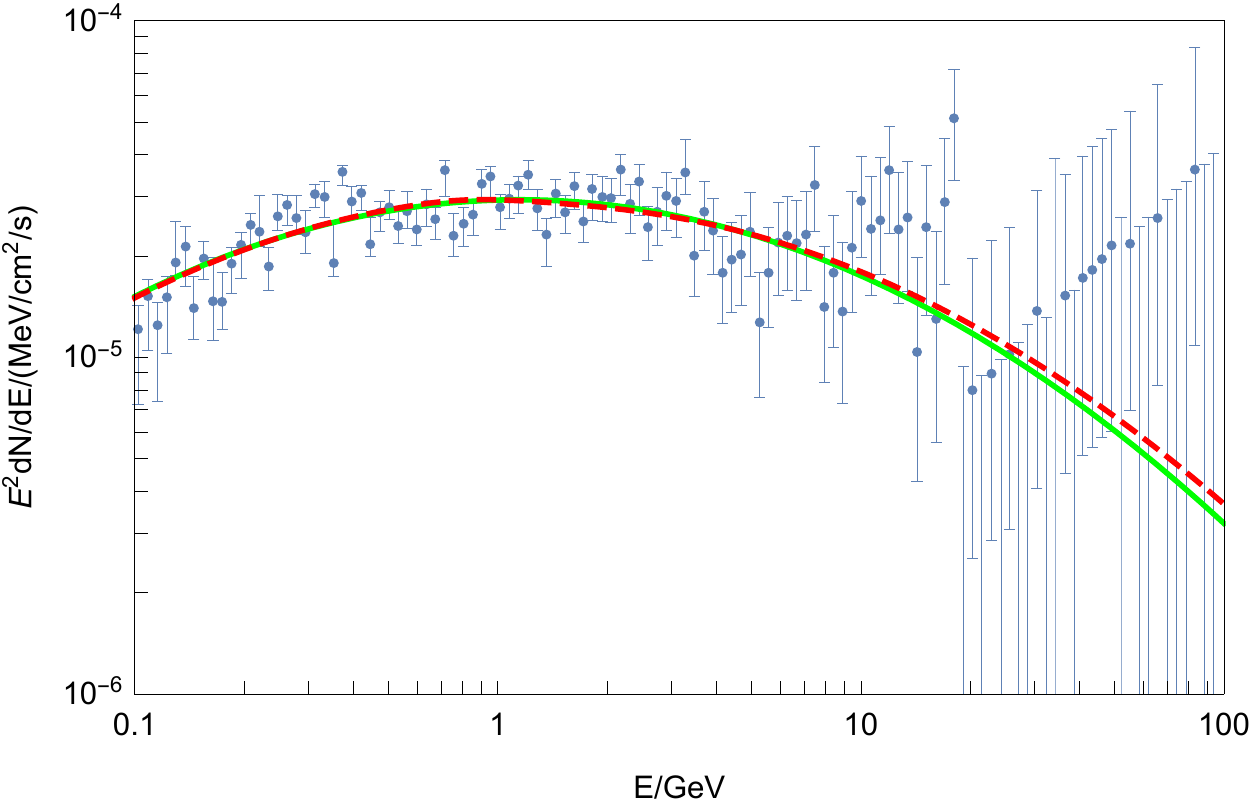}}
\caption{\label{fig:spectrum}
Fermi-LAT spectrum of NGC~1275 \cite{1603.06978} (data points) together
with two best-fit spectra: with (red dashed line) and without (green full
line) ALPs with $m=1$~neV, $g=10^{-11}$~GeV$^{-1}$.}
\end{figure}

For the same benchmark values
of $m=10^{-9}$~eV and $g=10^{-11}$~GeV$^{-1}$, the two best-fit spectra
are shown in Fig.~\ref{fig:spectrum}
together with the data. The fit quality is determined by $\chi^{2}
\approx 115.9$ for the fit without ALPs and $\chi^{2}\approx 115.6$ for
the fit with ALPs, for 114 degrees of freedom. We see that both fits are
perfect, and the effects of ALPs with the benchmark parameters cannot be
excluded for the regular field model used here, while they are excluded
at the 95\% CL for the purely turbulent field model in
Ref.~\cite{1603.06978}.

\begin{figure}
\centerline{\includegraphics[width=0.95\columnwidth]{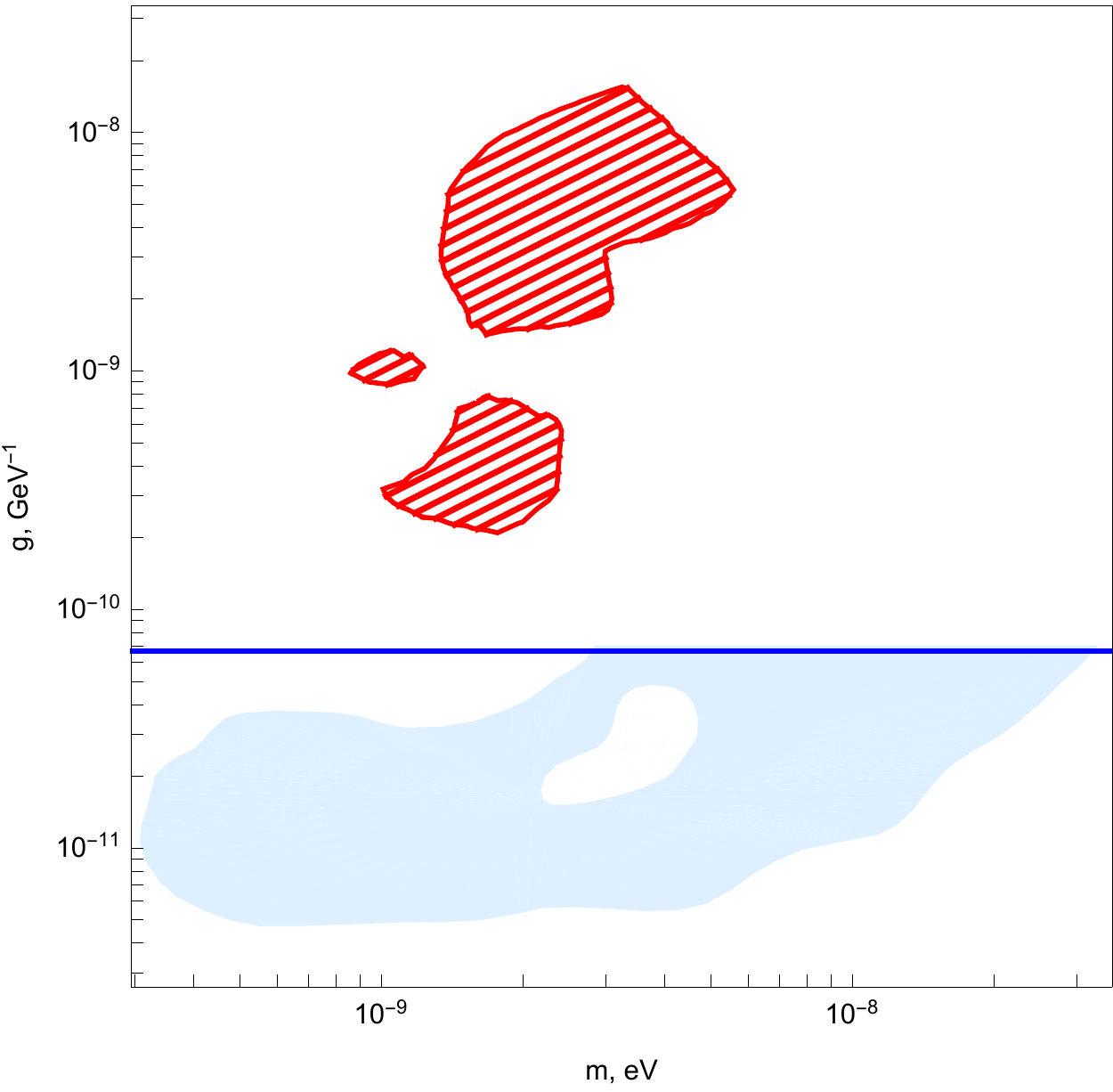}}
\caption{\label{fig:explot}
Exclusion regions (95\% CL) on the plane of ALP parameters $(m,g)$ for
purely regular (this work, hatched) and purely turbulent
(Ref.~\cite{1603.06978}, shaded) magnetic fields around NGC~1275. The blue
horizontal line represents the CAST~\cite{CAST} 95\% CL upper limit on
$g$. Ref.~\cite{1603.06978} does not extend the exclusion above the CAST
limit. For the regular field considered here, all exclusion regions are
above the CAST limit. For a realistic combination of regular and
turbulent fields, the exclusion is somewhere in between, but one needs to
know magnetic fields around NGC~1275 in more detail to find where exactly.}
\end{figure}

This fitting procedure was repeated for various pairs of ALP parameters
$(m,g)$. The 95\% exclusion contour in the $(m,g)$ plane was determined
from the chi-squared distribution with 114 degrees of freedom,
$\chi^{2}_{95} \approx 139.9$. This contour is shown in
Fig.~\ref{fig:explot}
together with the exclusion region
obtained in Ref.~\cite{1603.06978} for purely turbulent fields.
The difference between purely regular and purely turbulent cases is
dramatic. Since the actual magnetic field around NGC~1275 is an unknown
mixture of regular and turbulent components, this difference gives an
estimate of theoretical uncertainties of the constraints.

\red
The main physical reason behind the difference between the resulting
constraints on ALP parameters obtained in
this section and in Ref.~\cite{1603.06978}  is the use of coherent
versus turbulent fields. While the photon-ALP conversion probability
builds up for large-scale coherent fields, reaching eventually the
maximal-mixing regime, it is not the probability which determines the
strength of the constraint: instead, it is the presence and the amplitude
of ``wiggles'' around $\sim 1$~GeV, where the spectrum is well measured,
cf.\ Fig.~\ref{fig:pconv}. In a multi-scale turbulent field, the
irregularities are enhanced as compared to the regular-field case, because
for certain energies the oscillation length fits the field coherence scale
and, for this energy, the conversion probability is locally enhanced,
hence a wiggle in the spectrum appears (see e.g.\ Ref.~\cite{1311.3148}).
Therefore, while the value of the amplitude of the central field used in
Ref.~\cite{1603.06978}, 10~$\mu$G, is slightly larger than the central
value of $8.3~\mu$G used here, this cannot be the reason for dramatically
different constraints. We note in passing that while our toy model is
normalized to reproduce the observed Faraday rotation
measure~\cite{Taylor-FRM}, it is very hard to obtain its large observed
value with purely turbulent fields since positive and negative
contributions cancel each other in the corresponding integral.

Another difference with the analysis of
Ref.~\cite{1603.06978} is in the event classes used. We base our toy
research on the published spectrum and energy resolution, available for
the EDISP3 class only. This high-quality set, by construction, contains
1/4 of the data. These data, with their good energy resolution,
dominate constraints on ``wiggles'' in the spectrum; however the full set
of data was used in Ref.~\cite{1603.06978}, which gives an additional
caveat in the direct comparison of the two results.

Finally, one more difference is that Ref.~\cite{1603.06978}, as one may
conclude from a detailed description~\cite{1406.5972}, used the
domain-like approximation for the magnetic field while we solved the full
equations for smooth fields numerically. It has been pointed out (see
e.g.\ Ref.~\cite{1804.09443}) that the domain approximation, like the one
used in Ref.~\cite{1406.5972}, may introduce certain biases in the
probability calculation. We checked by an explicit full calculation that
this effect arises at high energies and hence has a limited influence on
the spectral fitting because of large statistical errors of the measured
spectrum in this energy range.

 \black

\section{Conclusions}
\label{sec:concl}
Magnetic fields regular on large scales are expected to be present around
giant active galaxies, in particular around those which are used to
constrain ALP parameters from the lack of irregularities in their spectra.
Particular examples are NGC~1275 and PKS~2155$-$304, for which
high-quality gamma-ray and X-ray spectra are available. In this paper, we
revisit constraints from gamma-ray observations of NGC~1275 which assumed
purely turbulent magnetic fields around this giant radio galaxy. We note
that an X-ray cavity observed around the galaxy suggests the presence of
ordered magnetic fields and use a model of the cavity field to study how
the ALP constraints are changed in the presence of regular fields.
Assuming that the observed Faraday rotation measurements are fully
explained by the cavity field, we obtain constraints much weaker than
those obtained in Ref.~\cite{1603.06978} for a purely turbulent field
model, see Fig.~\ref{fig:explot}. In reality, the field is expected to
include both components, and actual constraints would lay somewhere in
between, but unfortunately the lack of magnetic-field measurements in the
Perseus cluster currently prevents one from disentangling the two
contributions.

\red The key message of our paper -- that a good knowledge of the magnetic
field is a necessary prerequisite of a study aming to constrain ALP
parameters from the search of spectral irregularities, -- remains valid
for observations in other energy bands, not only in gamma rays. For
instance, for the X-ray energy range, additional complications arise
because of non-negligible effects of the electron density, which drive the
photon-ALP conversion to the outer parts of the cluster, where the
field is even less constrained than in the center.

\black Similar arguments are applicable to other sources, in
particular to PKS~2155$-$304.
\red Previous studies~\cite{1311.3148} assumed that the group, in which
this blazar resides~\cite{PKSgroup1}, possesses magnetic fields similar to
much larger clusters\footnote{\red Coma ($\sim 1000$ members, $M_{\rm
vir}\sim 1.7 \times 10^{15} M_{\odot}$, $R_{\rm vir}\sim 2.9$~Mpc
\cite{MNRAS-343-401}) and Hydra ($\sim 160$ members, $M_{\rm vir}\sim
2\times 10^{14} M_{\odot}$, $R_{\rm vir}\sim 1.6$~Mpc
\cite{Hydra-cluster}) clusters. The group containing PKS has the virial
radius of $R_{\rm vir} \sim 0.22$~Mpc, virial mass $M_{\rm vir}\sim 1.5
\times 10^{13} M_{\odot}$ and only 12 identified group
members~\cite{1510.01779}. Only in a few cases (not including this one),
magnetic fields in such small groups have been determined observationally,
and these fields were regular (see e.g.\ Ref.~\cite{1701.05962} and
references therein), not turbulent.}. No Faraday rotation measurements are
available for this group, so Ref.~\cite{1311.3148} and follow-up studies
made use of this theoretical model of the magnetic field without any
direct observational data.
\black

While new
observations of the same targets with sensitive instruments like CTA
\cite{CTA} or ATHENA$+$ \cite{ATHENA} will reduce statistical
uncertainties
in
the spectra and are certainly welcome for many reasons, uncertainties in
the magnetic-field models should first be removed by complementary
observations. In view of results of Ref.~\cite{FRM-map}, Faraday rotation
mapping of the Perseus cluster looks a promising way to obtain better
constraints on the magnetic field in the region where the lobes of
NGC~1275 interact with the intracluster matter, see e.g.\ Ref.\
\cite{Bonafede}. In addition, sources embedded in magnetic fields better
studied observationally may be invoked for the search of irregularities,
for instance M~87~\cite{1703.07354} and Galactic sources~\cite{1801.01646,
1801.08813, 1804.07186}. For certain values of ALP parameters, the
transition to the regime of strong mixing in the Galactic magnetic field
may be observed or constrained also for extragalactic sources, cf.\
Ref.~\cite{1805.04388}. Note that the relevant part of the ALP parameter
space will be explored by coming experiments, a purely laboratory
instrument ALPS-IIc \cite{ALPS} and helioscopes TASTE \cite{TASTE} and
IAXO \cite{IAXO-CDR}.

\section*{Acknowledgements}
The authors are indebted to Mikhail Kuznetsov, Manuel Meyer, Maxim
Pshirkov, Grigory Rubtsov and Dmitri Semikoz for discussions
and useful comments on the manuscript. ST acknowledges discussions
with Igor Garcia Irastorza and Manuel Meyer at the initial stage of this
work. We are especially thankful to the anonymous reviewer for very useful
comments which improved the paper. This work was supported by the Russian
Science Foundation, grant 18-12-00258.




\begin{thebibliography}{78}


\bibitem{Ringwald-rev}
 J.~Jaeckel and A.~Ringwald,
``The Low-Energy Frontier of Particle Physics,''
  Ann.\ Rev.\ Nucl.\ Part.\ Sci.\  {\bf 60} (2010) 405,
  doi:10.1146/annurev.nucl.012809.104433
  [arXiv:1002.0329 [hep-ph]].

\bibitem{RaffStod}
G.~Raffelt and L.~Stodolsky,
``Mixing of the Photon with Low Mass Particles,''
Phys.\ Rev.\  D {\bf 37}, 1237 (1988),
doi:10.1103/PhysRevD.37.1237  .

\bibitem{Losses}
M.~Giannotti {\it et al.},
``Cool WISPs for stellar cooling excesses,''
  JCAP {\bf 1605} (2016)  057,
doi:10.1088/1475-7516/2016/05/057
  [arXiv:1512.08108 [astro-ph.HE]].

\bibitem{HBstars}
A.~Ayala  {\it et al.},
``Revisiting the bound on axion-photon coupling from Globular Clusters,''
  Phys.\ Rev.\ Lett.\  {\bf 113} (2014)  191302,
doi:10.1103/PhysRevLett.113.191302
  [arXiv:1406.6053 [astro-ph.SR]].

\bibitem{CAST}
V.~Anastassopoulos {\it et al.} [CAST Collaboration],
  ``New CAST Limit on the Axion-Photon Interaction,
  Nature Phys.\  {\bf 13} (2017) 584,
  doi:10.1038/nphys4109
  [arXiv:1705.02290 [hep-ex]].

\bibitem{hardening}
S.~Archambault {\it et al.} [VERITAS and Fermi-LAT Collaborations],
``Deep Broadband Observations of the Distant Gamma-ray Blazar PKS
1424+240,''
Astrophys.\ J.\  {\bf 785} (2014) L16,
doi:10.1088/2041-8205/785/1/L16
[arXiv:1403.4308  [astro-ph.HE]].

\bibitem{3c279}
 E.~Aliu {\it et al.} [MAGIC Collaboration],
``Very-High-Energy Gamma Rays from a Distant Quasar: How Transparent Is
the Universe?,''
Science {\bf 320} (2008) 1752,
doi:10.1126/science.1157087
  [arXiv:0807.2822 [astro-ph]].

\bibitem{2photons}
Y.~T.~Tanaka
{\it et al.},
``Fermi Large Area Telescope detection of two very-high-energy
($E>100$~GeV) gamma-ray photons from the $z = 1.1$ blazar PKS~0426$-$380,''
Astrophys.\  J.\ {\bf 777} (2013) L18,
doi:10.1088/2041-8205/777/1/L18
[arXiv:1308.0595].

\bibitem{HornsMeyer}
 D.~Horns and M.~Meyer,
``Indications for a pair-production anomaly from the propagation of VHE
gamma-rays,''
JCAP {\bf 1202} (2012) 033,
doi:10.1088/1475-7516/2012/02/033
[arXiv:1201.4711 [astro-ph.CO]].

\bibitem{gamma}
 G.~I.~Rubtsov and S.~V.~Troitsky,
 ``Breaks in gamma-ray spectra of distant blazars and transparency of the
Universe,''
JETP Lett.\  {\bf 100} (2014)  355 [Pis'ma ZhETF  {\bf 100}
  (2014) 397],
doi:10.1134/S0021364014180088
[arXiv:1406.0239 [astro-ph.HE]].

\bibitem{Csaba}
C.~Csaki {\it et al.},
  ``Super GZK photons from photon axion mixing,''
  JCAP {\bf 0305} (2003) 005,
doi:10.1088/1475-7516/2003/05/005
  [hep-ph/0302030].

\bibitem{DARMA}
 A.~De Angelis, M.~Roncadelli and O.~Mansutti,
``Evidence for a new light spin-zero boson from cosmological gamma-ray
propagation?,''
Phys.\ Rev.\ D {\bf 76} (2007) 121301,
doi:10.1103/PhysRevD.76.121301
[arXiv:0707.4312  [astro-ph]].

\bibitem{Serpico}
M.~Simet, D.~Hooper and P.~D.~Serpico,
``The Milky Way as a Kiloparsec-Scale Axionscope,''
  Phys.\ Rev.\ D {\bf 77} (2008) 063001,
doi:10.1103/PhysRevD.77.063001
  [arXiv:0712.2825 [astro-ph]].

\bibitem{FRT}
 M.~Fairbairn, T.~Rashba and S.~V.~Troitsky,
``Photon-axion mixing and ultra-high-energy cosmic rays from BL Lac type
objects - Shining light through the Universe,''
Phys.\ Rev.\ D {\bf 84}  (2011) 125019,
doi:10.1103/PhysRevD.84.125019
[arXiv:0901.4085 [astro-ph.HE]].

\bibitem{1207.0776clusters}
D.~Horns
{\it et al.},
``Hardening of TeV gamma spectrum of AGNs in galaxy clusters by
conversions of photons into axion-like particles,''
Phys.\ Rev.\ D {\bf 86} (2012) 075024,
doi: 10.1103/PhysRevD.86.075024
[arXiv:1207.0776 [astro-ph.HE]].

\bibitem{Meyer-evidence}
 M.~Meyer, D.~Horns and M.~Raue,
``First lower limits on the photon-axion-like particle coupling from
very high energy gamma-ray observations,''
Phys.\ Rev.\ D {\bf 87} (2013) 035027,
doi:10.1103/PhysRevD.87.035027
[arXiv:1302.1208 [astro-ph.HE]].

\bibitem{Galanti-spindex}
G.~Galanti {\it et al.},
``Axion-like particles explain the unphysical redshift-dependence of AGN
gamma-ray spectra,''
arXiv:1503.04436 [astro-ph.HE].

\bibitem{ST-rev}
 S.~V.~Troitsky,
``Axion-like particles and the propagation of gamma rays over astronomical
distances,''
JETP Lett.\  {\bf 105} (2017) 55,
doi:10.1134/S0021364017010052
[arXiv:1612.01864 [astro-ph.HE]].

\bibitem{gamma2}
A.~Korochkin, G.~Rubtsov and S.~Troitsky,
``Distance-dependent hardenings in gamma-ray blazar spectra corrected for
the absorption on the extragalactic background light,''
arXiv:1810.03443 [astro-ph.HE].

\bibitem{astro-ph/0410501}
 L.~Ostman and E.~Mortsell,
``Limiting the dimming of distant Type Ia supernovae,''
  JCAP {\bf 0502} (2005) 005,
  doi:10.1088/1475-7516/2005/02/005
  [astro-ph/0410501].

\bibitem{0806.0411}
D.~Chelouche {\it et al.},
``Spectral Signatures of Photon-Particle Oscillations from Celestial
Objects,''
Astrophys.\ J.\ Suppl.\  {\bf 180} (2009) 1,
  doi:10.1088/0067-0049/180/1/1
[arXiv:0806.0411 [astro-ph]].

\bibitem{1205.6428}
D.~Wouters and P.~Brun,
``Irregularity in gamma ray source spectra as a signature of axionlike
particles,''
Phys.\ Rev.\ D {\bf 86} (2012) 043005,
  doi:10.1103/PhysRevD.86.043005
[arXiv:1205.6428 [astro-ph.HE]].

\bibitem{1305.2114}
G.~Galanti and M.~Roncadelli,
``Comment on ``Irregularity in gamma ray source spectra as a signature of
axion-like particles'',''
  arXiv:1305.2114 [astro-ph.HE].

\bibitem{1406.5972}
M.~Meyer, D.~Montanino and J.~Conrad,
``On detecting oscillations of gamma rays into axion-like particles in
turbulent and coherent magnetic fields,''
  JCAP {\bf 1409} (2014) 003,
  doi:10.1088/1475-7516/2014/09/003
  [arXiv:1406.5972 [astro-ph.HE]].

\bibitem{1804.09443}
G.~Galanti and M.~Roncadelli,
``Behavior of axionlike particles in smoothed out domainlike magnetic
fields,''
  Phys.\ Rev.\ D {\bf 98} (2018) 043018,
  doi:10.1103/PhysRevD.98.043018
  [arXiv:1804.09443 [astro-ph.HE]].

\bibitem{1811.03548}
G.~Galanti {\it et al.},
``Blazar VHE spectral alterations induced by photon-ALP oscillations,''
  doi:10.1093/mnras/stz1144
  [arXiv:1811.03548 [astro-ph.HE]].

\bibitem{1311.3148}
A.~Abramowski {\it et al.} [H.E.S.S. Collaboration],
``Constraints on axionlike particles with H.E.S.S. from the irregularity of
the PKS 2155$-$304 energy spectrum,''
  Phys.\ Rev.\ D {\bf 88} (2013) 102003,
  doi:10.1103/PhysRevD.88.102003
  [arXiv:1311.3148 [astro-ph.HE]].

\bibitem{1802.08420}
C.~Zhang {\it et al.},
``New bounds on axionlike particles from the Fermi Large Area Telescope
observation of PKS 2155$-$304,''
  Phys.\ Rev.\ D {\bf 97} (2018) 063009,
  doi:10.1103/PhysRevD.97.063009
  [arXiv:1802.08420 [hep-ph]].

\bibitem{1906.00357}
J.~Bu and Y.~P.~Li,
``Constraints on axionlike particles with different magnetic field models
from the PKS 2155-304 energy spectrum,''
arXiv:1906.00357 [astro-ph.HE].

\bibitem{1603.06978}
M.~Ajello {\it et al.} [Fermi-LAT Collaboration],
``Search for spectral irregularities due to photon-axionlike-particle
oscillations with the Fermi Large Area Telescope,''
Phys.\ Rev.\ Lett.\  {\bf 116} (2016) 161101,
doi:10.1103/PhysRevLett.116.161101
[arXiv:1603.06978 [astro-ph.HE]].

\bibitem{1805.04388}
D.~Malyshev  {\it et al.},
``Improved limit on axion-like particles from $\gamma$-ray data on Perseus
cluster,''
  arXiv:1805.04388 [astro-ph.HE].

\bibitem{1801.01646}
Z.~Q.~Xia {\it et al.},
``Searching for spectral oscillations due to photon-axionlike particle
conversion using the Fermi-LAT observations of bright supernova remnants,''
  Phys.\ Rev.\ D {\bf 97} (2018) 063003,
  doi:10.1103/PhysRevD.97.063003
  [arXiv:1801.01646 [astro-ph.HE]].

\bibitem{1801.08813}
J.~Majumdar, F.~Calore and D.~Horns,
  ``Search for gamma-ray spectral modulations in Galactic pulsars,''
  JCAP {\bf 1804} (2018) 048,
  doi:10.1088/1475-7516/2018/04/048
  [arXiv:1801.08813 [hep-ph]].

\bibitem{1804.07186}
Y.~F.~Liang {\it et al.},
``Constraints on axion-like particle properties with very high energy
gamma-ray observations of Galactic sources,''
  arXiv:1804.07186 [hep-ph].

\bibitem{1304.0989}
D.~Wouters and P.~Brun,
``Constraints on axion-like particles from X-ray observations of the Hydra
galaxy cluster,''
Astrophys.\ J.\  {\bf 772} (2013) 44,
  doi:10.1088/0004-637X/772/1/44
[arXiv:1304.0989 [astro-ph.HE]].

\bibitem{1605.01043}
M.~Berg {\it et al.},
``Constraints on Axion-Like Particles from X-ray Observations of NGC1275,''
  Astrophys.\ J.\  {\bf 847} (2017) 101,
  doi:10.3847/1538-4357/aa8b16
  [arXiv:1605.01043 [astro-ph.HE]].

\bibitem{1712.08313}
L.~Chen and J.~P.~Conlon,
``Constraints on massive axion-like particles from X-ray observations of
NGC 1275,''
  Mon.\ Not.\ Roy.\ Astron.\ Soc.\  {\bf 479} (2018) 2243,
  doi:10.1093/mnras/sty1591
  [arXiv:1712.08313 [astro-ph.HE]].

\bibitem{1907.05475}
C.~S.~Reynolds {\it et al.},
``Astrophysical limits on very light axion-like particles from Chandra
grating spectroscopy of NGC 1275,''
arXiv:1907.05475 [hep-ph].

\bibitem{1703.07354}
M.~C.~D.~Marsh {\it et al.},
  ``A New Bound on Axion-Like Particles,''
  JCAP {\bf 1712} (2017) 036,
  doi:10.1088/1475-7516/2017/12/036
  [arXiv:1703.07354 [hep-ph]].

\bibitem{1704.05256}
J.~P.~Conlon {\it et al.},
``Constraints on axion-like particles from non-observation of spectral
modulations for X-ray point sources,''
JCAP {\bf 1707} (2017) 005,
  doi:10.1088/1475-7516/2017/07/005
[arXiv:1704.05256 [astro-ph.HE]].

\bibitem{PDG}
M.~Tanabashi {\it et al.} [Particle Data Group],
``Review of Particle Physics,''
  Phys.\ Rev.\ D {\bf 98} (2018) 030001,
  doi:10.1103/PhysRevD.98.030001 .

\bibitem{a-p/0204443}
P.~Alexander,
``On the interaction of FR-II radio sources with the intracluster medium,''
  Mon.\ Not.\ Roy.\ Astron.\ Soc.\  {\bf 335} (2002) 610,
  doi:10.1046/j.1365-8711.2002.05638.x
  [astro-ph/0204443].

\bibitem{1108.0430}
M.~Huarte-Espinosa, M.~Krause and P.~Alexander,
``Interaction of Fanaroff-Riley class II radio jets with a randomly
magnetised intra-cluster medium,''
  Mon.\ Not.\ Roy.\ Astron.\ Soc.\  {\bf 418} (2011) 1621,
  doi:10.1111/j.1365-2966.2011.19545.x
  [arXiv:1108.0430 [astro-ph.CO]].

\bibitem{1203.4582}
D.~Guidetti {\it et al.},
 ``The magnetized medium around the radio galaxy B2 0755$+$37: an
interaction with the intra-group gas,''
Mon.\ Not.\ Roy.\ Astron.\ Soc.\   {\bf 423} (2012) 1335,
doi:10.1111/j.1365-2966.2012.20961.x
[arXiv:1203.4582 [astro-ph.CO]].

\bibitem{1811.06266}
J.~K.~Banfield  {\it et al.},
``Faraday rotation study of NGC 612 (PKS 0131$-$36): a hybrid radio source
and its magnetized circumgalactic environment,''
Mon.\ Not.\ Roy.\ Astron.\ Soc.\  {\bf 482} (2019) 5250,
doi:10.1093/mnras/sty3108
[arXiv:1811.06266 [astro-ph.HE]].

\bibitem{1812.07900}
B. Adebahr, M. Brienza, R. Morganti,
``Polarised structures in the radio lobes of B2 0258$+$35 - Evidence of
magnetic draping?'',
Astron.\ Astrophys.\ {\bf 622} (2019) A209,
doi:10.1051/0004-6361/201833988
[arXiv:1812.07900 [astro-ph.GA]].

\bibitem{1101.1807}
D.~Guidetti {\it et al.},
``Ordered magnetic fields around radio galaxies:
evidence for interaction with the environment,''
  Mon.\ Not.\ Roy.\ Astron.\ Soc.\  {\bf 413} (2011) 2525,
  doi:10.1111/j.1365-2966.2011.18321.x
  [arXiv:1101.1807 [astro-ph.CO]].

\bibitem{1612.01764}
M.~Kierdorf {\it et al.},
``Relics in galaxy clusters at high radio
frequencies'',
Astron.\ Astrophys.\ {\bf 600} (2017) A18,
doi:10.1051/0004-6361/201629570
[arXiv:1612.01764 [astro-ph.GA]].

\bibitem{1205.1919}
L.~Feretti {\it et al.},
``Clusters of galaxies: observational properties of the diffuse
radio emission,''
  Astron.\ Astrophys.\ Rev.\  {\bf 20} (2012) 54,
  doi:10.1007/s00159-012-0054-z
  [arXiv:1205.1919 [astro-ph.CO]].

\bibitem{Han-ARAA55(17)111}
J.~L.~Han,
``Observing interstellar and intergalactic magnetic fields,''
Ann.\ Rev.\ Astron.\ Astrophys.\ {\bf 55} (2017) 111,
doi:10.1146/annurev-astro-091916-055221 .

\bibitem{Taylor-FRM}
 G.~B.~Taylor {\it et al.},
 ``Magnetic fields in the center of the Perseus cluster,''
  Mon.\ Not.\ Roy.\ Astron.\ Soc.\  {\bf 368} (2006) 1500,
  doi:10.1111/j.1365-2966.2006.10244.x
  [astro-ph/0602622].

\bibitem{FRM-map}
A.~G.~de Bruyn and M.~A.~Brentjens,
  ``Diffuse polarized emission associated with the Perseus cluster,''
  Astron.\ Astrophys.\  {\bf 441} (2005) 931,
  doi:10.1051/0004-6361:20052992
  [astro-ph/0507351].

\bibitem{MNRAS264(93)L25}
H.~Boehringer {\it et al.},
``A ROSAT HRI study of the interaction of the X-ray emitting gas and radio
lobers of NGC 1275,''
Mon.\ Not.\ Roy.\ Astron.\ Soc.\  {\bf 264} (1993) L25,
doi:10.1093/mnras/264.1.L25 .

\bibitem{a-p/0207290}
R.~W.~Schmidt, A.~C.~Fabian and J.~S.~Sanders,
  ``Chandra temperature and metallicity maps of the Perseus cluster core,''
  Mon.\ Not.\ Roy.\ Astron.\ Soc.\  {\bf 337} (2002) 71,
  doi:10.1046/j.1365-8711.2002.05804.x
  [astro-ph/0207290].

\bibitem{a-p/0503318}
J.~S.~Sanders, A.~C.~Fabian and R.~J.~H.~Dunn,
  ``Non-thermal X-rays, a high abundance ridge and fossil bubbles in the
core of the Perseus cluster of galaxies,''
  Mon.\ Not.\ Roy.\ Astron.\ Soc.\  {\bf 360} (2005) 133,
  doi:10.1111/j.1365-2966.2005.09016.x
  [astro-ph/0503318].

\bibitem{1701.03791}
M.~Gendron-Marsolais {\it et al.},
  ``Deep 230-470 MHz VLA observations of the mini-halo in the Perseus
cluster,''
  Mon.\ Not.\ Roy.\ Astron.\ Soc.\  {\bf 469} (2017) 3872,
  doi:10.1093/mnras/stx1042
  [arXiv:1701.03791 [astro-ph.CO]].

\bibitem{1008.5353}
K.~N.~Gourgouliatos, J.~Braithwaite and M.~Lyutikov,
  ``Structure of magnetic fields in intracluster cavities,''
  Mon.\ Not.\ Roy.\ Astron.\ Soc.\  {\bf 409} (2010) 1660,
  doi:10.1111/j.1365-2966.2010.17410.x
  [arXiv:1008.5353 [astro-ph.HE]].

\bibitem{1011.0030}
H.~Xu {\it et al.},
 ``Evolution and distribution of magnetic fields from AGNs in galaxy
clusters. I. The effect of injection energy and redshift,''
Astrophys.\   J.\  {\bf 725} (2010) 2152,
 doi:10.1088/0004-637X/725/2/2152
  [arXiv:1011.0030 [astro-ph.CO]].

\bibitem{1107.2599}
H.~Xu {\it et al.},
``Evolution and distribution of magnetic fields from AGNs in galaxy
clusters.
II. The effects of cluster size and dynamical state,''
  Astrophys.\ J.\  {\bf 739} (2011) 77,
  doi:10.1088/0004-637X/739/2/77
  [arXiv:1107.2599 [astro-ph.CO]].

\bibitem{1108.3344}
P.~M.~Sutter {\it et al.},
 ``An examination of magnetized outflows from active galactic nuclei in
galaxy clusters,''
  Mon.\ Not.\ Roy.\ Astron.\ Soc.\  {\bf 419} (2012) 2293,
  doi:10.1111/j.1365-2966.2011.19875.x
  [arXiv:1108.3344 [astro-ph.CO]].

\bibitem{Churazov}
 E.~Churazov {\it et al.},
 ``XMM-Newton observations of the Perseus cluster I: the temperature and
surface brightness structure,''
  Astrophys.\ J.\  {\bf 590} (2003) 225,
  doi:10.1086/374923
  [astro-ph/0301482].

\bibitem{1810.07380}
Y.~Ichinohe {\it et al.},
 ``Substructures associated with the sloshing cold front in the Perseus
cluster,''
  Mon.\ Not.\ Roy.\ Astron.\ Soc.\  {\bf 483} (2019) 1744,
  doi:10.1093/mnras/sty3257
  [arXiv:1810.07380 [astro-ph.HE]].

\bibitem{1506.06429}
N.~Werner {\it et al.},
``Deep Chandra observation and numerical studies of the nearest cluster
cold front in the sky,''
  Mon.\ Not.\ Roy.\ Astron.\ Soc.\  {\bf 455} (2016) 846,
  doi:10.1093/mnras/stv2358
  [arXiv:1506.06429 [astro-ph.CO]].

\bibitem{PKSgroup1}
R.~Falomo, J.~E.~Pesce and A.~Treves,
  ``The Environment of the BL Lac object PKS 2155$-$304,''
  Astrophys.\ J.\  {\bf 411} (1993) L63,
  doi:10.1086/186913

\bibitem{1510.01779}
 E.~P.~Farina {\it et al.},
  ``The cluster-scale environment of PKS 2155$-$304,''
  Mon.\ Not.\ Roy.\ Astron.\ Soc.\  {\bf 455} (2016) 618,
  doi:10.1093/mnras/stv2277
  [arXiv:1510.01779 [astro-ph.GA]].

\bibitem{1602.03099}
M.~L.~Ahnen {\it et al.} [MAGIC Collaboration],
``Deep observation of the NGC 1275 region with MAGIC: search of diffuse
$\gamma$-ray emission from cosmic rays in the Perseus cluster,''
  Astron.\ Astrophys.\  {\bf 589} (2016) A33,
doi:10.1051/0004-6361/201527846 [arXiv:1602.03099 [astro-ph.HE]].

\bibitem{AN-327-545}
M.~A.~Brentjens and A.~G.~de Bruyn,
``RM-synthesis of the Perseus cluster,''
Astron.\ Nachr.\ {\bf 327} (2006) 545,
doi:10.1002/asna.200610584 .

\bibitem{PshirkovGMF}
 M.~S.~Pshirkov {\it et al.},
``Deriving global structure of the Galactic Magnetic Field from Faraday
Rotation Measures of extragalactic sources,''
  Astrophys.\ J.\  {\bf 738} (2011) 192,
  doi:10.1088/0004-637X/738/2/192
  [arXiv:1103.0814 [astro-ph.GA]].

\bibitem{physics/0401042}
R.~Barlow,
``Asymmetric errors,''
  eConf C {\bf 030908} (2003) WEMT002
  [physics/0401042 [physics.data-an]].

\bibitem{1701.05962}
J.~F.~Kaczmarek {\it et al.},
``Detection of a Coherent Magnetic Field in the Magellanic Bridge through
Faraday Rotation,''
Mon.\ Not.\ Roy.\ Astron.\ Soc.\  {\bf 467} (2017) 1776,
  doi:10.1093/mnras/stx206
  [arXiv:1701.05962 [astro-ph.GA]].

\bibitem{MNRAS-343-401}
E.~L.~Lokas and G.~A.~Mamon,
``Dark matter distribution in the Coma cluster from galaxy kinematics:
Breaking the mass - anisotropy degeneracy,''
  Mon.\ Not.\ Roy.\ Astron.\ Soc.\  {\bf 343} (2003) 401,
  doi:10.1046/j.1365-8711.2003.06684.x
  [astro-ph/0302461].

\bibitem{Hydra-cluster}
T.~Sato {\it et al.},
``Suzaku observations of the Hydra A cluster out to the virial radius,''
  Publ.\ Astron.\ Soc.\ Jap.\  {\bf 64} (2012) 95,
  doi:10.1093/pasj/64.5.95
  [arXiv:1203.1700 [astro-ph.CO]].

\bibitem{CTA}
M.~Actis {\it et al.} [CTA Consortium],
``Design concepts for the Cherenkov Telescope Array CTA: An advanced
facility for ground-based high-energy gamma-ray astronomy,''
  Exper.\ Astron.\  {\bf 32} (2011) 193,
  doi:10.1007/s10686-011-9247-0
  [arXiv:1008.3703 [astro-ph.IM]].

\bibitem{ATHENA}
  K.~Nandra {\it et al.},
``The Hot and Energetic Universe: A White Paper presenting the science
theme motivating the Athena$+$ mission,''
  arXiv:1306.2307 [astro-ph.HE].

\bibitem{Bonafede}
A.~Bonafede {\it et al.} [SKA magnetism working group],
``Unravelling the origin of large-scale magnetic fields in galaxy clusters
and beyond through Faraday Rotation Measures with the SKA,''
  PoS AASKA {\bf 14} (2015) 095,
  doi:10.22323/1.215.0095
  [arXiv:1501.00321 [astro-ph.CO]].

\bibitem{ALPS}
R.~Bahre {\it et al.},
``Any light particle search II -Technical Design Report,''
  JINST {\bf 8} (2013) T09001,
doi:10.1088/1748-0221/8/09/T09001
  [arXiv:1302.5647 [physics.ins-det]].

\bibitem{TASTE}
V.~Anastassopoulos {\it et al.},
``Towards a medium-scale axion helioscope and haloscope,''
  JINST {\bf 12} (2017) P11019,
  doi:10.1088/1748-0221/12/11/P11019
  [arXiv:1706.09378 [hep-ph]].

\bibitem{IAXO-CDR}
E.~Armengaud {\it et al.},
``Conceptual Design of the International Axion Observatory (IAXO),''
  JINST {\bf 9} (2014) T05002,
  doi:10.1088/1748-0221/9/05/T05002
  [arXiv:1401.3233 [physics.ins-det]].

\end{thebibliography}


\end{document}